\documentclass[doublecol,revision]{epl2} 

\usepackage{amssymb}
\usepackage{amsmath}
\usepackage{amsfonts}
\usepackage{graphicx}
\usepackage{mathrsfs}
\usepackage{bm}
\usepackage{color}
\usepackage{mathbbol}
\usepackage{appendix}

\title{On the origin of power-laws in equilibrium}

\shorttitle{On the origin of power-laws in equilibrium} 

\author{Michele Campisi\inst{1} \and Fei Zhan\inst{2} \and Peter H\"anggi\inst{1}}
\shortauthor{M. Campisi \etal}

\institute{                    
  \inst{1} Institute of Physics, University of Augsburg,
  Universit\"atsstr. 1, D-86159 Augsburg, Germany\\
  \inst{2} International Center for Quantum Materials, Peking University, 100871, Beijing,
China}

\abstract{
A  particle in the attractive Coulomb field has an interesting property:
its specific heat is constant and negative. We show, both analytically and numerically, that
 when a classical Hamiltonian system stays in 
weak contact with one such negative specific heat object, its statistics
conforms to a fat-tailed power-law distribution with power index  
given by $C/k_B-1$, where $k_B$ is Boltzmann constant and $C$ is the heat capacity.}

\begin{document}

\maketitle

\section{Introduction}

In 1968 Lynden-Bell and Wood \cite{Lynden-Bell68MNRAS138} pointed out 
an interesting fact: self-gravitating systems have negative 
specific heats. 
As Lynden-Bell later on explained \cite{LyndenBell99PHYSA293} this fact seemed quite natural
to astronomers, who know for example that when a star gains energy it expands and cools down \cite{Antonov95IAUS113};
but it appeared incongruous, if not completely wrong, to statistical mechanists, who learn from textbooks that specific heats are necessarily positive. The paradox was resolved by Thirring
\cite{Thirring70ZPA235}: While it is true that any system \emph{in weak contact with a thermal bath}, hence 
characterized by the canonical distribution, necessarily has a positive specific heat,
isolated systems, characterized by the microcanonical distribution, may well have 
negative specific heats. Since then, many models showing microcanonical 
negative specific heats have been reported,
see, e.g., \cite{Thirring03PRL91,Barre01PRL87,Borges02PHYSA305,Hilbert06PRE74,Dunkel06PHYSA370,Casetti10JSM,Kastner10PRL104}, their statistical mechanics has been discussed \cite{Padmanabhan90PR285,Chavanis06IJMPB20},
and negative specific heats have been experimentally measured in thermally isolated small atomic clusters
\cite{Schmidt01PRL86,Gobet02PRL89}. Recently it has also been pointed out that if a system is
in \emph{strong} coupling with its environment, likewise it may display negative specific heat 
\cite{Hanggi06APP37,Hanggi08NJP10,Zitko09PRB79,Campisi09JPA42,Campisi10CP375,Dattagupta10PRE81,Hasegawa11JMP52,Ingold12EPJB85}.

Inspired by Ref. \cite{Almeida01PHYSA424}, here
we consider an ordinary system $S$ with Hamiltonian $H_S(\mathbf{x},\mathbf{p})$ and 
\emph{weakly} couple it via an interaction energy $h(\mathbf{x},\mathbf{X})$, to a second system with Hamiltonian 
$H_C(\mathbf{X},\mathbf{P})$, possessing a \emph{constant} negative microcanonical heat capacity $C<0$, i.e.,
\begin{equation}
H(\mathbf{x},\mathbf{p},\mathbf{X},\mathbf{P})=H_S(\mathbf{x},\mathbf{p})+H_C(\mathbf{X},\mathbf{P})+h(\mathbf{x},\mathbf{X})\,.
\label{eq:H-3/2}
\end{equation}
For the sake of clarity we recall that
the heat capacity $C$ is defined as the derivative of a system's energy 
with respect to the temperature, $C=\partial E/\partial T$, while the specific heat $c$ is defined as the heat capacity per unit mass.

The archetypical example of a system with 
constant negative microcanonical heat capacity is that
of a single particle in  the gravitational (or the attractive Coulomb) force field, for which $C=-3/2$ \cite{LyndenBell99PHYSA293}. Throughout the paper, temperature
is expressed in units of energy. In these units $k_B$, Boltzmann's constant, is dimensionless and equal to $1$, 
and the heat capacity $C=\partial E/\partial T$ is dimensionless as well. 

Our main result is that, provided the \emph{total} system samples the microcanonical ensemble, 
the system $S$ samples the power-law distribution 
\begin{eqnarray}
 p(\mathbf{x},\mathbf{p})=\frac{[H_S(\mathbf{x},\mathbf{p})-E_{\text{tot}}]^{C-1}}
 {\int \mathrm{d}\mathbf{x}\mathrm{d}\mathbf{p} [H_S(\mathbf{x},\mathbf{p})-E_{\text{tot}}]^{C-1}}, \quad C<0 .
\label{eq:power-law}
\end{eqnarray}
where $E_\text{tot}$ is the (conserved) energy of the total system. To express this result in Eq. (\ref{eq:power-law}) 
in the usual set of units, where the temperature is measured in Kelvin, and both the heat capacity and $k_B$ 
have the dimension of Joule/Kelvin, one should replace $C$ in Eq. (\ref{eq:power-law}) 
with the dimensionless ratio $C/k_B$.

\begin{figure}
\begin{center}
\includegraphics[width=.3\textwidth]{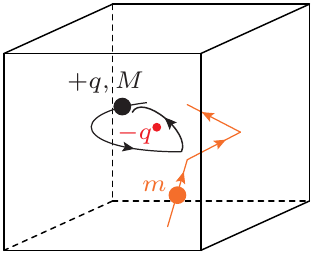}
\end{center}
\caption{\label{fig:ske3d}(Color online) Illustrative example of our main results, Eq. (\ref{eq:power-law}). A neutral particle (orange) moves freely within a box, making elastic collisions with the box walls and a charged particle (black) immersed in the attractive Coulomb field generated by a fixed point charge (red). Because the heat capacity of the charged (black) particle is $C=-3/2$, the velocity probability distribution function of the neutral (orange) particle conforms to a power law with power index $-3/2-1=-5/2$, Eq. (\ref{eq:power5/2}). }
\end{figure}
We illustrate this result in Fig. \ref{fig:ske3d}. 
A neutral particle, with Hamiltonian $H_S=\bm{p}^2/2m$, is confined into a box and
makes elastic collisions with a particle carrying a charge $q$ subject to the Coulomb field generated by a fixed charge $-q$. The charged particle has constant negative heat capacity $C=-3/2$, and its Hamiltonian reads:
\begin{equation}
H_{C=-3/2}(\mathbf{X},\mathbf{P})=  
\bm{P}^2/2M - \alpha/|\mathbf X|\, ,
\label{eq:H-3/2}
\end{equation}
where $\alpha =q^2/4\pi \varepsilon_0$
with $\varepsilon_0$ the dielectric permittivity of vacuum.
According to our main result, Eq. (\ref{eq:power-law}), since the charged particle has constant negative heat capacity $C=-3/2$, the velocity probability distribution function of the neutral particle is given by the 
following power law
\begin{equation}
p(\bm{p})=
\frac{(\bm{p}^2/2m-E_{\text{tot}})^{-5/2}}
{\int d\bm{p}(\bm{p}^2/2m-E_{\text{tot}})^{-5/2}}\, .
\label{eq:power5/2}
\end{equation}

\section{Derivation of the result in Eq. (\ref{eq:power-law})} Assuming (i) the microcanonical distribution
for the total system, and (ii) weak coupling $h$, the system $S$
marginal distribution $ p(\mathbf{x},\mathbf{p})$ can be written as \cite{Khinchin49Book,Campisi09PRE80,Campisi12PRL108}
\begin{eqnarray}
 p(\mathbf{x},\mathbf{p})=\frac{\Omega_ C[E_{\text{tot}}-H_S(\mathbf{x},
\mathbf{p})]}{
\Omega_\text{tot}(E_{\text{tot}})}\, ,
\label{eq:khinchin-formula}
\end{eqnarray}
where
\begin{equation}
 \Omega_C(E) = \int \mathrm{d}{\bm{X}}
\mathrm{d}\bm{P}\delta[E-H_C(\bm{X}, \bm{P})]
\label{eq:Omega-B}
\end{equation}
is the density of states of the negative heat capacity system whose energy is $E$, and
\begin{equation}
\Omega_\text{tot}(E_{\text{tot}}) = \int \mathrm{d}\bm{X}
\mathrm{d}\bm{P}\mathrm{d}\mathbf{x}\mathrm{d}\mathbf{p}\,
\delta[E_{\text{tot}}-H(\mathbf{x},\mathbf{p},\bm{X},\bm{P})]
\end{equation}
is the density of states of the compound system. Here $\delta$ denotes
Dirac's delta function. 

We note that if a classical system has 
a constant \emph{negative} microcanonical heat capacity $C<0$, then its density of states is of the form:
\begin{equation}
\Omega_C(E)\propto (-E)^{C-1}\, ,\quad C<0 \, .
\label{eq:statement1}
\end{equation}
For the derivation of this density of states see Appendix A.
In arriving at Eq. (\ref{eq:statement1}) we adopt  the convention to set the zero of the energy of constant negative heat capacity systems as the lowest energy corresponding to unbound trajectories. Thus the density of states is only defined for negative $E$'s, and it diverges for $E\geq 0$. 

Using Eq. (\ref{eq:statement1}) in (\ref{eq:khinchin-formula}) one arrives at the result in Eq. (\ref{eq:power-law}).
It is important to stress that in Eq. (\ref{eq:power-law}) the system $S$ energy has a lower bound, which is conventionally set to zero, and that the total energy $E_\text{tot}$ must be negative, thus ensuring that the energy  $E=E_\text{tot}-H_S$
 of the system with negative heat capacity is negative at all times. This is necessary in order that the
 $C$-system always stays on bounded trajectories and never escapes to infinity.
 
\section{Numerics}
To corroborate our main result, Eq. (\ref{eq:power-law}), we simulated the dynamics of the system depicted in Fig. \ref{fig:ske3d},
using a symplectic integrator \cite{HairerBook}.
We focussed on the probability density $\rho(E_S)$ of finding the neutral particle (our $S$ system)
at energy $E_S$ for a total (negative) simulation energy $E_\text{tot}$. According to Eq. (\ref{eq:power5/2}) this is given by
\begin{equation}
\rho(E_S)=\frac{[E_S-E_{\text{tot}}]^{{-5}/{2}}E_S^{1/2}}{\int_{0}^{\infty} d E_S [E_S-E_{\text{tot}}]^{{-5}/{2}}E_S^{1/2}}\, ,
\label{eq:energyPDF}
\end{equation}
where the term $E_S^{1/2}$ derives from the density of states $\Omega_S(E_S)\propto E_S^{3/2-1}$ of the neutral particle. Note that for large $E_S$, $\rho(E_S) \propto E_S^{-2}$.

Following Ref. \cite{Campisi12PRL108}, we chose
$h(\mathbf{x},\mathbf{X})=V_{LJ}(|\mathbf{x}-\mathbf{X}|)$, 
where $V_{LJ}$ is the truncated Leonard-Jones potential
\begin{equation}
V_{LJ}(r)=
 \left\{
  \begin{array}{ll}
0\, , &  |r| > 2^{1/6}\sigma \\
4\varepsilon \left[\left(\frac{\sigma}{r}\right)^{12}-\left(\frac{\sigma}{r}
\right)^ { 6 } \right ] +\varepsilon\, , & |r| < 2^{1/6}\sigma
 \end{array}
 \right. \, .
\label{eq:V(x)}
\end{equation}

We employ the same potential for the walls of the confining cubic box, in which the particle moves freely. In our simulation, we set $\varepsilon$, $\sigma$, and $M$ ($M$ is the mass of the particle carrying the charge $q$) as the units of energy, length, and mass respectively.

Following Ref. \cite{Campisi12PRL108}, in order to avoid the singularity of the attractive Coulomb field at $\mathbf X=0$, in the simulation we employ the Plummer potential \cite{plummer11mnras}:
\begin{equation}
\varphi_b(\mathbf{X})= \frac{-\alpha}{\sqrt{\bm{X}^2+b^2}}\, .
\label{eq:Plummer}
\end{equation}
Using this potential in place of the pure Coulombic potential $\varphi(\mathbf{X})= {-\alpha}/{|\mathbf X|}$, induces a deviation of the density of states
from the pure power-law form $\Omega_{-3/2}(E)=(-E)^{-5/2}$, resulting in 
a cut-off in the expected power law distribution $\rho(E_S)$. This cut-off moves to higher energy values as $b$ becomes smaller.

\begin{figure}
\begin{center}
\includegraphics[width=.46\textwidth]{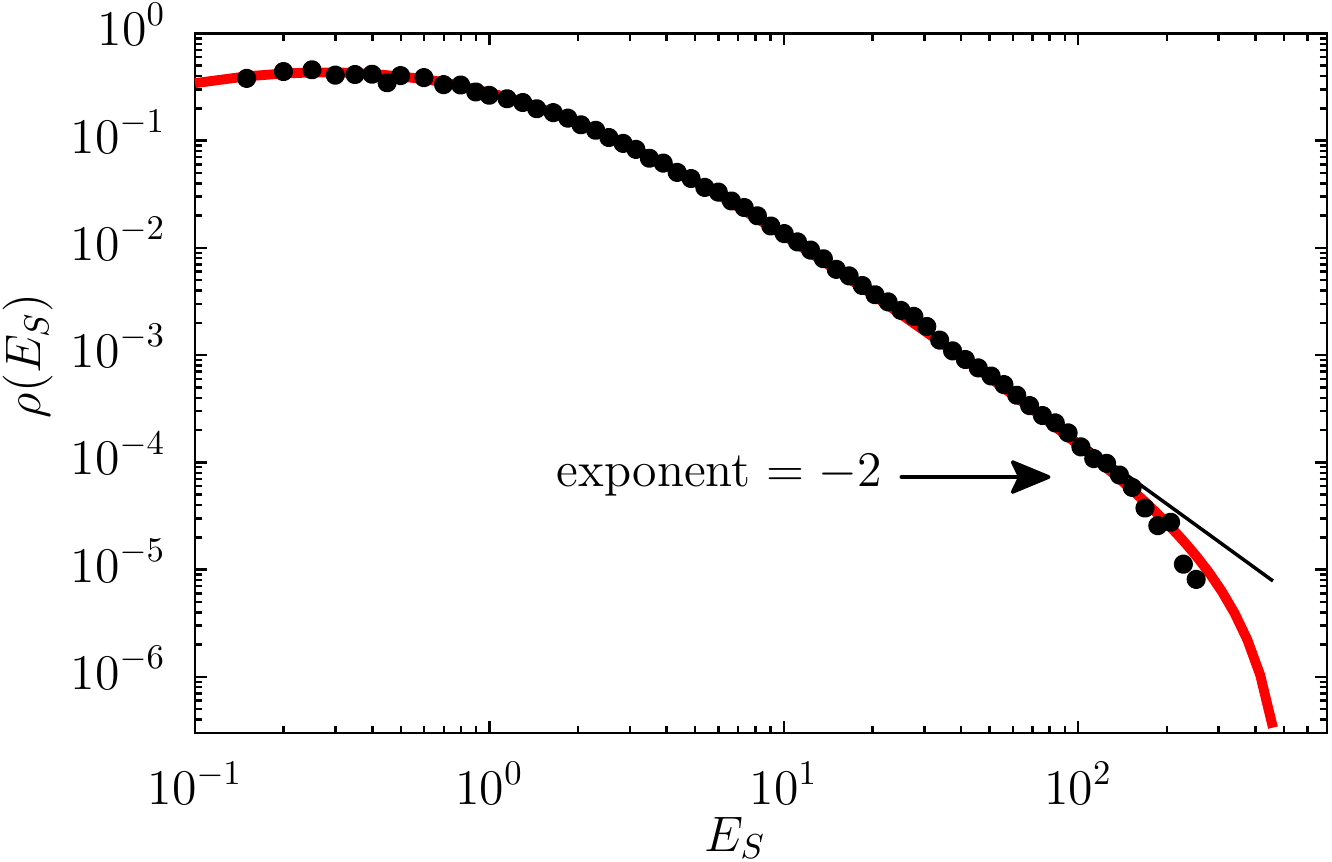}
\end{center}
\caption{\label{fig:power}(Color online) Energy probability density function $\rho(E_S)$ of a neutral particle performing Lennard-Jones collisions, Eq. (\ref{eq:V(x)}), with the walls of a confining box, and 
with a charged particle in the Plummer potential, Eq. (\ref{eq:Plummer}).
Black dots: numerical simulation. Red thick solid line: analytic calculation based on the Plummer potential, Eq. (\ref{eq:Plummer}). Black thin solid line: analytic calculation based on the pure gravitational potential $\varphi(\mathbf{X})= {-\alpha}/{|\mathbf X|}$, Eq. (\ref{eq:energyPDF}). Simulation parameters: $E_\text{tot}=-1$, $b=10^{-2}$, $L=10$, $\alpha=5$, and $m=\sqrt{3}$.}
\end{figure}

Fig. \ref{fig:power} displays the result of our simulation for $b=10^{-2}$, $E_\text{tot}=-1$,
a box of side length $L=10$, $\alpha=5$, and $m=\sqrt{3}$.
The computed energy distribution function for the neutral atom excellently agrees with the
expected power-law form in Eq. (\ref{eq:energyPDF}) over three decades.
At higher energies the distribution displays the expected cut-off due to the truncation in Eq. (\ref{eq:Plummer}). Note that when the neutral atom has high energy the charged particle
has low energy and stays close to the bottom of the Plummer potential, Eq. (\ref{eq:Plummer}).
The condition $E_\text{tot}<0$ ensures that the energy $E=E_\text{tot}-H_S$ of the charged particle is also negative, thus ensuring that its motion remains confined at all times.
With our choice of box side length $L=10$, and $E_\text{tot}=-1$, the confinement keeps the charged particle within the box at all times.

\subsection{Methods} The simulation was performed by means of an implicit Runge-Kutta symplectic integrator \cite{HairerBook}. We simulated the evolution 
of the whole system with a time step size $\Delta t=10^{-5}$ and total number of time steps $\mathcal{N}=1.28\times 10^{12}$. 
Note that the units of time are $\sqrt{M/\varepsilon}\sigma$. 
In order to calculate the statistics,
we divided the $E_S$-axis in $10^4$ intervals of size $\Delta E_S=0.0499$,
and counted how many times each interval was visited during the simulation.
We adopted a sampling time $\tau=2\times 10^5\Delta t$.
The resulting histogram was then normalized and divided by $\Delta E_S$ to 
give an estimate of the probability density. Finally, for better visualization in the log-log plot in Fig. \ref{fig:power},
the histogram was coarse-grained with a series of intervals with exponentially increasing size, viz., we took $\Delta E^{n+1}_S/\Delta E^{n}_S=1.1$, where $\Delta E^{n}_S$ is the size of the $n$th interval and $\Delta E^{0}_S=\Delta E_S$.

\section{Comparison with the Finite Bath Statistics}

When $C>0$, one finds the expression $\Omega_C(E)\propto E^{C-1}$ for the densities of states, see Appendix A. This differs from Eq. (\ref{eq:statement1}) by a minus sign in front of $E$. Note that for constant positive heat capacity systems, whose energy is bounded from below, we adopt the convention that the zero of the energy corresponds to this lower bound. Accordingly, in this case one obtains for the probability density function the same expression as in Eq. (\ref{eq:power-law}), but with $H_S-E_\text{tot}$ replaced by $E_\text{tot}-H_S$:
\begin{eqnarray}
 p(\mathbf{x},\mathbf{p})=\frac{[E_{\text{tot}}-H_S(\mathbf{x},\mathbf{p})]^{C-1}}
 {\int \mathrm{d}\mathbf{x}\mathrm{d}\mathbf{p}[E_{\text{tot}}-H_S(\mathbf{x},\mathbf{p})]^{C-1}},\quad C>0 \, .
\label{eq:power-law2}
\end{eqnarray}
The statistics in Eq. (\ref{eq:power-law2}) is sometimes referred to as the ``finite bath statistics'' \cite{Campisi09PRE80}, because it is induced by the weak coupling of the $S$ system with a system with a finite \emph{positive} heat capacity, like, for instance a finite collection of $N$ free particles, for which $C=3N/2$. As we will see below, this can also be achieved with one single particle in a properly chosen potential.  The finite bath statistics 
in Eq. (\ref{eq:power-law2}) interpolates between the microcanonical distribution $p(\mathbf{x},\mathbf{p})\propto \delta[E_S-H_S(\mathbf{x},\mathbf{p})]$ and the canonical distribution $p(\mathbf{x},\mathbf{p})\propto e^{-H_S(\mathbf{x},\mathbf{p})/T}$. These two limiting cases are obtained when $C$ approaches $0$ and $+\infty$ respectively \cite{Campisi07PLA366,Campisi07PHYSA385,Campisi09PRE80}.
Note that when $C>0$,  $E_{\text{tot}}-H_S$ in Eq. (\ref{eq:power-law2}), is non-negative, as a consequence of the fact that the system $S$ energy may not overcome the total (non-negative) energy $E_\text{tot}$.
While the negative $C$ distributions in Eq. (\ref{eq:power-law}) have fat power-law tails,
the positive $C$ distributions in Eq. (\ref{eq:power-law2}) exhibits sharp cut-offs at $H_S=E_\text{tot}$.

\section{Implementing constant heat capacity systems with single particles}
In three dimensions, constant heat capacity systems
can be obtained with a single particle governed by the  
Hamiltonian:
\begin{equation}
H_{C}(\mathbf{X},\mathbf{P})=   \bm{P}^2/2M + g C |\mathbf{X}/L|^{6/(2C-3)}\, ,
\label{eq:HC}
\end{equation}
with constants $g,L>0$ with units of energy and length respectively, and $C\notin [0,3/2]$. As shown with explicit calculations in Appendix B,
the associated density of states is proportional to $E^{C-1}$ for $C>3/2$, and 
is proportional to $(-E)^{C-1}$ for $C<0$.
Accordingly, when a system with Hamiltonian $H_S$ bounded from below stays in weak contact with
one such constant heat capacity system, its statistics conforms with either Eq. (\ref{eq:power-law}) or Eq. (\ref{eq:power-law2}), depending on whether $C<0$ or $C>3/2$, respectively.
The potential in (\ref{eq:HC}) includes many special cases of interest.
For $C$ approaching $3/2$ the potential approaches a box potential, namely the system is a particle in a box,
which knowingly has a constant heat capacity of $3/2$. For $C=3$, we have a harmonic oscillator. For $C=-3/2$ we have the $1/r$ case of Eq. (\ref{eq:H-3/2}). For $C \rightarrow \pm \infty$, the potential approaches the logarithmic form:
\begin{equation}
C |\mathbf{X}|^{6/(2C-3)} \simeq 3  \ln |\mathbf{X}| + C, \quad |C|\gg\ 1\, ,
\end{equation}
which, as we have shown recently \cite{Campisi12PRL108}, produces the Gibbs distribution. 
Thus, not only does the canonical ensemble emerge from the coupling to a system with positive infinite 
heat capacity, but also from the coupling to a system with \emph{negative} infinite 
heat capacity. The case $C=0$ is excluded from Eq. (\ref{eq:HC}) because it corresponds to an unbound free particle, 
whose density of states, accordingly, diverges for all energies.
For $0<C<3/2$, the exponent $6/(2C-3)$ is negative, hence one must chose $g<0$ to ensure that the potential is attractive.
The density of states diverges nonetheless because it involves a divergent integral, see Appendix B.

Similar expressions as in Eq. (\ref{eq:HC}), but with a different exponent for $|\mathbf{X}|$
can be found in spaces of dimensions other than three.

\section{Conclusions}
We have shown, both analytically and numerically, that weak coupling to 
a system with constant negative heat capacity leads to the emergence of 
velocity probability distributions with fat power-law tails. 
The power-law emerges because the $S$-system may withdraw indefinitely large amounts of
energy from the infinitely deep algebraic potential ``well'' that systems with constant negative heat capacity 
possess. 

Because of the fat tails it is possible that the negative $C$ distributions may not be normalizable.
This happens, for instance, when the $S$-system density of states goes asymptotically as $E_S^{\alpha}$,
with $\alpha>-C$. In this case the $S$-system ``wins'' over the $C$-system and the 
integral of $\Omega_C[E_\text{tot}-E_S]\Omega_S[E_S]$, which asymptotically goes like $\propto E_S^{C+\alpha-1}$, diverges.
For $\alpha >-C+1$,  $\Omega_C[E_\text{tot}-E_S]\Omega_S[E_S]$
asymptotically increase with increasing energies, while for $-C+1>\alpha>-C$ it vanishes for $E_S\rightarrow \infty$. As examples, consider the case of Fig. \ref{fig:ske3d}, but with many neutral particles. With two neutral particles
we have $-C+1=5/2>\alpha = 2 > -C=3/2$, that is the distribution would not be normalizable but would vanish
for large $E_S$. With three or more neutral particles we have $\alpha\geq7/2>5/2= -C+1$, and the distribution would be increasing for large $E_S$. Arguably the system never reaches equilibrium in such cases.

For our example of a neutral particle making collisions with a particle in the 
$1/r$ potential field, Fig. \ref{fig:ske3d} , the energy distribution has a power index $-2$, which is very close to the exponent observed in the energy distribution of cosmic rays \cite{Carlson12PT65,Letessier-Selvon11RMP83}.\footnote{Typically one looks at the cosmic ray differential spectrum \cite{Carlson12PT65,Letessier-Selvon11RMP83}, which is the derivative of the probability distribution with respect to energy. The cosmic ray spectrum has a power index $\simeq -3$. Accordingly, the energy probability distribution has a power index $\simeq -2$.}
This suggests that the power index $\simeq -2$ of the cosmic ray energy distribution 
might originate from multiple collisions of the cosmic particles with objects obeying 1/r potentials 
before they reach our instruments. We leave this as an open question.

\acknowledgments
The authors thank Armin Seibert, Stefano Ruffo and Lapo Casetti for comments.
This work was supported by the cluster of excellence
Nanosystems Initiative Munich (P.H.), the Volkswagen
Foundation project No. I/83902 (P.H., M.C.), and the DFG priority program SPP 1243 (P.H., F.Z.).

\section{Appendix A. Density of states for systems with constant heat capacity} 
A classical system has a constant negative microcanonical heat capacity $C$ if and only if its density of states is of the form:
\begin{equation}
\Omega_C(E)= f (-E)^{C-1}\, ,\quad C< 0 \, 
\label{eq:statement1b}
\end{equation}
with some constant $f>0$.

To see this consider
the microcanonical heat capacity:
\begin{equation}
C(E)= [\partial T(E)/\partial E]^{-1}\, ,
\label{eq:mcSpecHeat}
\end{equation}
where $T(E)$ is the microcanonical temperature.
For constant $C$ this implies 
\begin{equation}
T(E)=E/C+a\, ,
\end{equation} with some integration constant $a$.
Using the definition of microcanonical temperature 
\begin{equation}
T(E) = [\partial \ln \Phi_C(E)/\partial E]^{-1} \, ,
\label{eq:mcTemp}
\end{equation} 
one obtains, after a further integration
\begin{equation}
\Phi_C(E)= d [E/C+a]^{C}\ ,
\label{eq:PhiC}
\end{equation}
with some integration constant $d$.
Here 
\begin{equation}
\Phi_C(E)= \int \mathrm{d}\bm{X} \mathrm{d}\bm{P}\theta[E-H_C(\bm{X}, \bm{P})]\, .
\label{eq:PhiCgeneral}
\end{equation}
 is the phase space volume, and $\theta$ denotes Heaviside's step function.  Adopting the convention $\Phi_C(0)=\infty$, sets $a=0$, hence  $\Phi_C(E)= d [E/C]^{C}$. Recalling that $C<0$, this can be recast as  $\Phi_C(E) \propto (-E)^{C}$. Using the well-known relation 
 \begin{equation}
 \Omega_C(E)=\partial \Phi_C(E)/\partial E\, ,
 \label{eq:OmegaIsPartialPhi}
 \end{equation}
we obtain Eq. (\ref{eq:statement1b}).

To see that the reverse is also true, consider Eq. (\ref{eq:OmegaIsPartialPhi}).
We have, using Eq. (\ref{eq:statement1b}),
\begin{equation}
\Phi_C(E)= \int_{-\infty}^E \Omega_C(E') dE' = - f (-E)^C /C\, .
\end{equation}
Using (\ref{eq:mcTemp}), we obtain:
\begin{equation}
T(E)=E/C\, .
\end{equation}
From Eq. (\ref{eq:mcSpecHeat}) it then follows that the heat capacity is $C$.

Likewise, a classical system has a constant positive microcanonical heat capacity $C$ if and only if its density of states is of the form:
\begin{equation}
\Omega_C(E)= f E^{C-1}\, ,\quad C> 0
\label{eq:statement1c}
\end{equation}
with some constant $f>0$.

To see this the argument proceeds identically to the argument above,
with the only difference that  in the case $C>0$ the constant $a$ is set to zero as a consequence of
the convention $\Phi_C(0)=0$.

The argument supporting the reverse implication also proceeds identically as above,
with the only difference that now the phase volume reads $\Phi_C(E)=  f E^C/C$.

\section{Appendix B. Density of states for the Hamiltonians in Eq. (\ref{eq:HC})}
We begin with the case $C<0$, implying that the exponent
\begin{equation}
\gamma = 6/(2C-3)
\end{equation}
lies in the interval $-2<\gamma<0$. Using Eq. (\ref{eq:PhiCgeneral}) with Eq. (\ref{eq:HC}), recalling that $E,C<0$, the phase volume reads,
after integration over $\mathrm{d}{\bm{P}}$
\begin{align}
\Phi_C(E)&= \frac{(2\pi M)^{3/2}}{\Gamma(5/2)}\int \mathrm{d}\bm{X}[E-g C |\mathbf{X}|^\gamma]^{3/2} \\ \nonumber
&= 4\pi\frac{(2\pi M)^{3/2}}{\Gamma(5/2)}\int_0^{(E/gC)^{1/\gamma}} \mathrm{d}\rho \rho^2 [E-g C \rho^\gamma]^{3/2} \\ \nonumber
&= 4\pi\frac{(2\pi M)^{3/2}}{\Gamma(5/2)}\frac{(-E)^{3/2+3/\gamma}}{(-gC)^{3/\gamma}}\int_0^{1} \mathrm{d}y y^2 
[y^\gamma-1]^{3/2} \, ,
\end{align}
where we switched to spherical coordinates in the second line and we performed the change of variables 
$y=\rho(gC/E)^{1/\gamma}$ in the third line. The limit of integration $(E/gC)^{1/\gamma}$ represents the turning point of the 
closed orbit of negative energy $E$.
In the interval $-2<\gamma<0$ the integral converges and can be expressed in terms of the gamma function $\Gamma(x)$, as:
\begin{equation}
\int_0^{1} \mathrm{d}y y^2 [y^\gamma-1]^{3/2} =\frac{\sqrt{\pi}}{4}\frac{\Gamma(-C)}{\Gamma(3/2-C)}\, .
\end{equation}

Note that $3/2+3/\gamma=C$. Hence $\Phi_C\propto (-E)^C$. 
Accordingly, using Eq. (\ref{eq:OmegaIsPartialPhi}), we find that
the density of states is of the form $\Omega_C\propto (-E)^{C-1}$.

For $0<C<3/2$ it is $\gamma<-2$, hence one must now chose $g<0$ in order that the potential is 
attractive. One then arrives at the same expression as for the case $C<0$. In this case however the integral 
$\int_0^{1} \mathrm{d}y y^2 [y^\gamma-1]^{3/2} $ diverges.

For $C>3/2$ it is $\gamma >0$. One can proceed as above, recalling that now $E,C>0$, to obtain
\begin{align}
\Phi_C(E)&= 4\pi\frac{(2\pi M)^{3/2}}{\Gamma(5/2)}\int_0^{(E/gC)^{1/\gamma}} \mathrm{d}\rho \rho^2 [E-g C \rho^\gamma]^{3/2}  \nonumber \\
&= 4\pi\frac{(2\pi M)^{3/2}}{\Gamma(5/2)}\frac{\int_0^{1} \mathrm{d}y y^2 
[1-y^\gamma]^{3/2} 
}{(gC)^{3/\gamma}}E^{C}\end{align}
The integral in the numerator is well behaved for all $\gamma>0$, and can be expressed as:
\begin{equation}
\int_0^{1} \mathrm{d}y y^2 [1-y^\gamma]^{3/2} =\frac{\sqrt{\pi}}{4}\frac{\Gamma(C-1/2)}{\Gamma(C+1)}\, .
\end{equation}
Using  Eq. (\ref{eq:OmegaIsPartialPhi}), we find that the density of states is of the form $\Omega_C\propto E^{C-1}$.

\end{document}